\documentclass[12pt]{article}
\usepackage[utf8]{inputenc}
\usepackage[T1]{fontenc}
\usepackage{amsmath,amssymb, amsfonts}
\usepackage{hyperref}
\usepackage{icomma}
\usepackage[utf8]{inputenc}
\usepackage[T1]{fontenc}
\usepackage{color}
\usepackage{graphicx}
\usepackage{hyperref}
\usepackage{titlesec}
\usepackage{lmodern}
\usepackage{units}
\usepackage{marvosym}
\usepackage[usenames,dvipsnames,svgnames,table]{xcolor}
\usepackage{cite}

\newcommand{\G}{\Gamma}

\newcommand{\beq}{\begin{equation}}
\newcommand{\eeq}{\end{equation}}

\begin{document}
\title{The trouble with twisting (2,0) theory
 }
 \author{Louise Anderson and Hampus Linander}
 \date{}
\maketitle
\begin{center}

\vspace*{-1em}
Department of Fundamental Physics\\
Chalmers University of Technology\\
S-412 96 G\"oteborg, Sweden\\[3mm]
{\tt louise.anderson@chalmers.se, linander@chalmers.se}
\end{center}

\vspace{1em}
\begin{abstract}
 We consider a twisted version of the abelian $(2,0)$ theory placed upon a Lorenzian six-manifold with a product structure, $M_6=C \times M_4 $. This is done by an investigation of the free tensor multiplet on the level of equations of motion, where the problem of its formulation in Euclidean signature is circumvented by letting the time-like direction lie in the two-manifold $C$ and performing a topological twist along $M_4$ alone. A compactification on $C$ is shown to be necessary to enable the possibility of finding a topological field theory. The hypothetical twist along a  Euclidean $C$ is argued to amount to the correct choice of linear combination of the two supercharges scalar on $M_4$. This procedure is expected and conjectured to result in a topological field theory, but we arrive at the surprising conclusion that this twisted theory contains no $Q$-exact and covariantly conserved stress tensor unless $M_4$ has vanishing curvature. This is to our knowledge a phenomenon which has not been observed before in topological field theories. In the literature, the setup of the twisting used here has been suggested as the origin of the conjectured AGT-correspondence, and our hope is that this work may somehow contribute to the understanding of it.
\end{abstract}

\newpage

\tableofcontents
\newpage

\section{Introduction}

This work is an investigation of the topological twisting of the  $(2,0)$ theory which has been suggested to be relevant in the explanation of the origin of the AGT-conjecture. Herein, the simpler model of the free tensor multiplet is considered, and we find that the resulting twisted theory exhibits some curious, undesirable properties. The most severe of these is the lack of any satisfactory formulation of a stress tensor. This surprising result will be clear eventually, but let us first start at the very beginning.

The theory known as $(2,0)$ theory \cite{Witten:1995, Witten:ct46} is a six-dimensional superconformal theory that continue to resist attempts at unraveling its mysteries. One way to obtain information about the theory is to look at its different compactifications. For example, when compactified on a circle it gives rise to five-dimensional maximally supersymmetric Yang-Mills theory \cite{Seiberg:noteon16}. Recently a whole class of four-dimensional gauge theories have been constructed in this way by compactifying  $(2,0)$ theory on a two-dimensional Riemann surface with possible defects \cite{Gaiotto:N2dualities, Klemm:sclass, Witten:sclass}. This class of theories is sometimes referred to as ``class $\mathcal{S}$'' in the literature \cite{Gaiotto:2009, Gadde:2013}. The way these theories are obtained through compactification has led to a conjecture about the relation of certain objects in four-dimensional- and two-dimensional theories, the so-called AGT correspondence \cite{AGT}.

 More specifically, this correspondence states that the correlation functions in two-dimensional Liouville theory are related to the Nekrasov partition function \cite{Nekrasov:2002, Nekrasov:2003} of certain $\mathcal{N}=2$ superconformal gauge theories in four dimensions. 
One natural way to derive it \cite{AGT, Yagi:1, Yagi:2} would be to link it to a certain geometric setup in $(2,0)$ theory,  where the spacetime is taken to be a product of a two-dimensional- and a four-dimensional manifold. In such a setting, compactifications could either be carried out on the two- or on the four-manifold, after which one could search for protected quantities which have survived the compactification. A relation should then exist between the protected quantities of both compactifications.

However, one is here faced with the great challenge of a lack of any satisfactory definition of $(2,0)$ theory that would permit such detailed calculations. 
While this is indeed true for the full, interacting $(2,0)$ theory, this is not the whole story for the abelian version. Here, a classical formulation in terms of equations of motion exists. 

Moreover, it is important to notice that for a general background all supersymmetry will be broken and such a situation cannot be expected to shed any light on the AGT-correspondence. In order to preserve some supersymmetry, one must first perform a topological twisting \cite{Witten:1988}. In a case where the six-manifold has the product structure mentioned previously, i.e. $M_6= C \times M_4$, and $M_6$ is of Euclidean signature, (thus the holonomy groups of both $C$ and $M_4$ are compact), such a theory admits a unique twisting, which has been claimed \cite{Yagi:1} to be analogous to the Donaldson-Witten twist of four-dimensional $\mathcal{N}=2$ Yang-Mills theory \cite{Witten:1988}. 
 In the literature (see for example \cite{Yagi:1, Witten:2011}), it has been stated that the six-dimensional twist of would result in a theory which would be topological along $M_4$ and holomorphic along $C$ \cite{Silverstein:Witten}. Herein, the behaviour of the Lorenzian theory (especially along the four-manifold), is investigated explicitly by computing a stress tensor. 
 Our result seems to indicate that this Lorenzian twist may as conjectured coincide with the Donaldson-Witten twist on a flat background, however it does not seem to be true  in the general case. A more detailed discussion of this shall be presented towards the end of this work.

However, the elusive side of $(2,0)$ theory once again comes back to bite us here, since not even the abelian version of this theory has a satisfactory description on a Euclidean six-manifold, but rather only on a six-manifold with Minkowski signature. In such a situation, the holonomy group would be non-compact, and a topological twisting that results in a scalar supercharge cannot be performed. If the light-like direction is taken to lie in $C$, one may still obtain supercharges that are scalars on $M_4$ by a twisting procedure. One of these charges has properties that would make it scalar along $C$ as well, were we in the Euclidean scenario. In this work, this is the supercharge we will consider, and the behaviour of the theory under it is the subject of investigation. The final conclusion is that, on a general $M_4$, the stress tensor of the theory cannot be both $Q$-exact and conserved, and the theory is thus not topological in the traditional sense.

The outline of this work is as follows:
In section 2 we describe the twisting procedure giving rise to the supercharge that is scalar on $M_4$ and give a detailed description of the field content in this new, twisted theory. Section 3 deals with the equations of motion as well as the supersymmetry transformations of the twisted theory. In section 4, a stress tensor is computed in the flat case which is shown to have all desired properties. An attempt at generalising this to a general $M_4$ is made, and any $Q$-exact stress tensor is shown to not be covariantly conserved. It is also shown that no modifications to either equations of motion or supersymmetry variations may be done which would rectify these obstructions when $M_4$ is curved.

\section{The Twisting} \label{sec:twist}
We consider the free tensor multiplet of the  $(2,0)$ theory on a flat six-manifold $M_6$ , endowed with a product metric such that  $M_6 = C \times M_4$, with $C$ some two-manifold and $M_4$ some four-manifold. Throughout this work, light-cone coordinates $\{+,-\}$ on $C$, and indices $\mu, \nu \in \{1,2,3,4\}$ denoting directions along $M_4$, will be used. When needed, $M, N \in \{0,1,2,3,4,5\}$ will denote indices in six dimensions.

The tensor multiplet \cite{Howe:1983} contains a symplectic Majorana-Weyl spinor $\Psi$ transforming in the $\mathbf{4}$ of the $R$-symmetry group $Spin(5)_R$, a scalar $\Phi$ in the $\mathbf{5}$ of $Spin(5)_R$ and a self-dual three-form $H_{M N P}$. This section will deal with the decomposition of these representations under the twist, and the next section will provide a detailed dictionary for reinterpreting these in terms of the field content of the twisted theory.

If $M_6$ is of Euclidean signature, as previously mentioned, the theory admits a unique topological twisting.  This since the $R$-symmetry group $Spin(5)$ contains a subgroup  $SU(2)_R \times U(1)_R$, which also may be found as subgroups of the Lorentz group of $C \times M_4$: $U(1) \times SU(2)_l \times SU(2)_r$. The twisting procedure is carried out by defining $SU(2)' $ to be the diagonal subgroup of $SU(2)_r \times SU(2)_R$ and $U(1)' $ as the same in $U(1) \times U(1)_R$. By considering the theory under the group $U(1)' \times SU(2)_l \times SU(2)'$, one finds a single supercharge which is scalar hereunder, and thus the possibility of a topological field theory exists.

However, the lack of a satisfactory formulation of the free tensor multiplet of $(2,0)$ theory in Euclidean signature forces us to work in a situation where $C$ is of Minkowski signature instead, with the correspondingly non-compact Lorentz group $Spin(1,1)$. There will thus be no way to embed this into $U(1)_R$, and hence it is not possible to perform a twisting along the two-manifold $C$ as in the above case. $M_4$ is however still of Euclidean signature, hence the twisting along these directions will not have been affected. This will be described in greater detail below.

In table \ref{tab:beforetwist} and  \ref{tab:aftertwist}, the representations of the fields and supersymmetry parameters before and after twisting along $M_4$ are shown. A more detailed explanation on how the six-dimensional field content should be translated to the fields of the twisted theory will as mentioned follow in the next section.

\begin{table}[h!]
  \begin{center}
    \begin{tabular}{ c l }
       &
      $  SU(2)_l \times SU(2)_r \times SU(2)_R \times U(1)_R  $ \\
      \hline
      $\Phi$&
      $  (  \mathbf{1},  \mathbf{1},   \mathbf{3}  )^0 \oplus (  \mathbf{1},  \mathbf{1},   \mathbf{1}  )^{\pm 1}  $ \\
      $\Psi$ &   
      $   (  \mathbf{2},  \mathbf{1},   \mathbf{2}  )^{\pm 1/2}  \oplus     (  \mathbf{1},  \mathbf{2},   \mathbf{2}  )^{\pm 1/2}    $ \\
      $H $ &
      $(  \mathbf{3},  \mathbf{1},   \mathbf{1}  ) ^0 \oplus   (  \mathbf{1},  \mathbf{3},   \mathbf{1}  ) ^0 \oplus ( \mathbf{2},  \mathbf{2},   \mathbf{1})^0 $ \\
      $\varepsilon$&
      $   (  \mathbf{2},  \mathbf{1},   \mathbf{2}  )^{\pm 1/2}  \oplus     (  \mathbf{1},  \mathbf{2},   \mathbf{2}  )^{\pm 1/2}    $ 
    \end{tabular}
  \end{center}
  \caption{Representations before twisting.}
  \label{tab:beforetwist}
\end{table}
\begin{table}[h!]
  \begin{center}
    \begin{tabular}{ c l l }
       & $ SU(2)_l \times SU(2)'  \times U(1)_R $ & Twisted fields \\
      \hline
      $\Phi$ & $  (  \mathbf{1},    \mathbf{3}  )^0 \oplus (  \mathbf{1},  \mathbf{1}  )^{\pm 1}  $ &
      $E_{\mu \nu}, \bar{\sigma}, \sigma  $ \\
      $\Psi$ & $   (  \mathbf{2},   \mathbf{2}  )^{\pm 1/2}  \oplus     (  \mathbf{1},  \mathbf{3},  )^{\pm 1/2}   \oplus     (  \mathbf{1},  \mathbf{1}  )^{\pm 1/2}    $&
      $\psi_\mu, \tilde{\psi}_\mu, \chi_{\mu \nu} ,\tilde{\chi}_{\mu \nu}, \eta ,  \tilde{\eta}$\\
      $H$ & $(  \mathbf{3},  \mathbf{1} ) ^0 \oplus   (  \mathbf{1},  \mathbf{3}) ^0 \oplus ( \mathbf{2},  \mathbf{2})^0 $ &
      $F^-_{\mu \nu}, F^+_{\mu \nu}, A_\mu$ \\
      $\varepsilon$ & $   (  \mathbf{2},   \mathbf{2}  )^{\pm 1/2}  \oplus     (  \mathbf{1},  \mathbf{3}  )^{\pm 1/2}   \oplus     (  \mathbf{1},  \mathbf{1} )^{\pm 1/2}    $ &
      $\dots, (\bar{\varepsilon}), \varepsilon$\\
    \end{tabular}
  \end{center}
  \caption{Representations after twisting.}
  \label{tab:aftertwist}
\end{table}

The superscripts indicate the charge under $U(1)_R$. For clarity, it should here be pointed out that the representations for the fermions and for the supersymmetry parameters  differ in their chirality on $C$ (which is not shown in table \ref{tab:beforetwist} and \ref{tab:aftertwist}).

\emph{If} we were in Euclidean signature, all of these new fields would also have charges under the $U(1)$ which would then be the Lorentz group of $C$. In the second step of the twisting procedure previously described, these charges would combine with the charges under $U(1)_R$. The charge under the new diagonal subgroup $U(1)'$ would then be given by the sum of these two charges. Hence the supercharge that would become scalar under such a twist would be the one with $U(1)_R$-charge of $-1/2$ whose parameter shall be denoted by $\varepsilon$. The other supersymmetry scalar on $M_4$, with $U(1)_R$-charge of $+1/2$, is denoted by $\bar{\varepsilon}$. That $\varepsilon$ is the parameter of interest can be seen by studying table \ref{tab:euclideantwist} where the representations after the four-twist in the Euclidean scenario is written down.
\begin{table}[h!]
\begin{center}
\begin{tabular}{c l }

&
 $U(1) \times SU(2)_l \times SU(2)'  \times U(1)_R
$ \\
\hline
$\Phi$ &
   $  (  \mathbf{1},    \mathbf{3}  )^0_0 \oplus (  \mathbf{1},  \mathbf{1}  )^{\pm 1}_0  $ \\
 $\Psi$ 
 &
$   (  \mathbf{2},   \mathbf{2}  )^{\pm 1/2}_{1/2}  \oplus     (  \mathbf{1},  \mathbf{3} )^{\pm 1/2}_{-1/2}    \oplus     (  \mathbf{1},  \mathbf{1}  )^{\pm 1/2} _{ -1/2}   $\\
$H$ 
&
  $(  \mathbf{3},  \mathbf{1} ) ^0_0 \oplus   (  \mathbf{1},  \mathbf{3}) ^0_0 \oplus ( \mathbf{2},  \mathbf{2})^0_0 $  \\
$\varepsilon$ 
 &
$   (  \mathbf{2},   \mathbf{2}  )^{\pm 1/2}_{ -1/2}  \oplus     (  \mathbf{1},  \mathbf{3} )^{\pm 1/2}_{ 1/2}          \oplus     (  \mathbf{1},  \mathbf{1}  )^{\pm 1/2} _{1/2}  $\\
\end{tabular}
\end{center}
\caption{Hypothetical Euclidean twist.}
\label{tab:euclideantwist}
\end{table}
The superscript here denotes the charge under the $U(1)_R$ whereas the subscripts denote the charges under the $U(1)$ Lorentz group of $C$.

One may choose some chiral, constant spinors $e^\pm$ to generate the two spinor representations for the fermions which are scalar on $M_4$,  namely $ (  \mathbf{1},  \mathbf{1}  )^{\pm 1/2}_{-1/2}$. (Again, the subscript denotes the charge under a hypothetical $U(1)$ Lorentz group of $C$, and is what distinguishes the two fermionic singlet representations  on $M_4$ from the ones of the supersymmetries.) In some cases, it is convenient to think of these two new base-spinors as complex linear combinations of constant symplectic Majorana-Weyl spinors, $e_1$ and $e_2$, such that $e^\pm=e_1 \pm i e_2$.

 The two spinors $e^\pm$ will as mentioned need to be chiral in the six-dimensional sense to generate the fermionic representations. $\G_+ e^\pm$ are then anti-chiral, constant spinors, which generate the $ (  \mathbf{1},  \mathbf{1})^{\pm 1/2}_{ +1/2}$ where the supersymmetry-charges that are of interest to us live. 

This allows for a parametrisation of the two supercharges which are scalar on $M_4$ in terms of some Grassmann parameters $u$ and $v$, together with a $\G$-matrix along $C$ to account for the six-dimensional chirality. These relations are given by:

\beq
\label{SUSY}
\varepsilon = v \G_+ e^- \hspace*{1.5cm}, \hspace*{1.5cm} \bar{\varepsilon}= u \G_+ e^+
,\eeq
where as repeatedly mentioned, the supersymmetry parameter that would become scalar on $C$ as well after a hypothetical further twist is $\varepsilon$.

\subsection{Details of reinterpreting the fields \label{sec:fields}}
The next order of business is to create a dictionary, translating the original field content of the six-dimensional free tensor multiplet  (table \ref{tab:beforetwist}) to the field content of the twisted theory (table \ref{tab:aftertwist}). 

\subsubsection*{Bosonic scalar}
Let the indices $i, j \in \{ 1,2,3\}$. One can then quite easily see that the self-dual two-form $E_{\mu \nu}$ of the twisted theory can be related to the first three components of the six-dimensional scalar field $\Phi$ as follows:
\begin{align}
\label{eq:phi_twist1}
E_{4 i}=& -\Phi_i \\ \nonumber
E_{ij}=& \epsilon_{i j k} \Phi^k.
\end{align}
Furthermore, the two last components of the six-dimensional scalar $\Phi$ are after twisting combined into a complex scalar $\sigma$:
\beq
\sigma=\frac{1}{\sqrt{2}}(\Phi_4-i \Phi_5)
.
\label{eq:phi_twist2}
\eeq

\subsubsection*{Bosonic three-form}
Reinterpreting the six-dimensional bosonic three-form in terms of the new, twisted fields is only slightly more complicated than the case of the scalars above. By using the fact that $H_{M N P}$ is self-dual, (with respect to the orientation and Riemannian structure on $M_6$), one may show that $ H_{+ \mu \nu}$ is a self-dual two-form in four dimensions, and $ H_{- \mu \nu}$ likewise is an anti-self-dual two-form on $M_4$ (all with respect to the orientation and Riemannian structure on $M_4$). This gives us a natural interpretation of the components of $H$ in terms of the twisted two-form $F$ as:
\begin{align}
  \label{three_form_twist}
  H_{+ \mu \nu}=& \frac{1}{2}\epsilon_{\mu \nu}{}^{\rho \sigma} H_{+\rho \sigma}  = F_{\mu \nu}^+\\ \nonumber
  H_{- \mu \nu}=& - \frac{1}{2}\epsilon_{\mu \nu}{}^{\rho \sigma} H_{-\rho \sigma}  = F_{\mu \nu}^-
  ,\end{align}
where $F_{\mu \nu}^\pm$ denotes the self-dual and anti-self-dual parts respectively.

Moreover, one may in a similar fashion interpret $H_{\mu \nu \rho}$ and $H_{+ - \sigma}$ in terms of the twisted one-form $A_\sigma$ and its dual as:
\begin{align}
H_{\mu \nu \rho}=  \epsilon_{\mu \nu \rho \sigma} H_{+ - }^\sigma=   \epsilon_{\mu \nu \rho \sigma}A^\sigma.
\label{three_form_twist2}
\end{align}

\subsubsection*{Fermionic fields}
$\Psi$ may be expanded in terms of the twisted fields $\eta$, $\psi$, \dots as follows:
\beq
\label{ferm_def}
\Psi= (\eta+\G_+\G_\mu \psi^\mu+\frac{1}{4}\G_\mu \G_\nu \chi^{\mu \nu})e^+   +    (\tilde{\eta}+\G_+\G_\mu \tilde{\psi}^\mu+\frac{1}{4} \G_\mu \G_\nu \tilde{\chi}^{\mu \nu})e^-
.\eeq
The terms in the above decomposition are precisely the twisted field content of the spinor field as given in table \ref{tab:aftertwist}.
By using how $e^\pm$ are related to symplectic Majorana-Weyl spinors, one can show that $\Psi$ indeed is a symplectic Majorana-Weyl spinor as well under the condition that the fields with- and without twiddles are related by complex conjugation. This also is consistent with the $U(1)_R$-charges of these different fields. 

However, in the case we wish to consider, namely the theory invariant under only the one supercharge that would become scalar in a Euclidean scenario, we must loosen these requirements on $\Psi$, since there is no such notion as a spinor being Majorana-Weyl on a six-dimensional Euclidean manifold. This means that the fields with- and without the twiddles will need to be considered as independent of one another non the less.

\subsection{Some useful relations}
To perform further calculations, we must first find ways to handle the $\G$-matrices which arise both in \eqref{ferm_def} when reinterpreting the fermionic spinor field in terms of the new, twisted ones, as well as in the expression for how the relevant supersymmetry parameter is written down in terms of our base spinors \eqref{SUSY}. In this section, some useful formulas for handling these are presented. 

The first, and maybe most important relation comes from the knowledge that our constant base spinors are singlets under all of the $SU(2)$'s after twisting, which gives us the relations
\begin{align}
\frac{1}{2}(\G_{4i}-\frac{1}{2} \epsilon_{i j k } \G^{j k}) e^\pm & = 0
 \\ \nonumber
\frac{1}{2}(\G_{4i}+\frac{1}{2} \epsilon_{i j k } \G^{j k})  e^\pm  +\frac{1}{2} \epsilon_{i j k } \G_R^{j k} e^\pm &  =0
.\end{align}
Here $\G$ denotes the $\G$-matrices of the Lorentz group, whereas $\G_R$ denotes the gamma matrices of the $R$-symmetry group. Again, the indices $\{ i,j,k \}$ take values in $\{1,2,3\}$. The top one of the above equations enforces that the $e^\pm$ are singlets under $SU(2)_l$, and the lower one reflects the same behaviour under $SU(2)'$. 

Furthermore, the charge under the $U(1)_R$ is known for the two spinors, and it is thus known how the generator of this group acts on them: 
\begin{align}
i\G_R^4 \G_R^5 e^\pm = \pm e^\pm
.\end{align}

A short calculation also shows that the action of one of these, say $\G_R^4$, corresponds to flipping the $U(1)_R$-charge and thus:
\begin{align}
\G_R^4 e^\pm = e^\mp
.\end{align}

Now we move on to relations involving the $\G$-matrices of the Lorentz group. The spinors are chiral in a six-dimensional manner, thus
\begin{align}
\G_0\G_1\G_2\G_3 \G_4 \G_5  e^\pm = e^\pm
.\end{align}
This may be reduced to chirality along $C$ and $M_4$ individually by studying how these representations decompose under the twisting procedure. If we let the six-dimensional indices be divided such that $\{ 0,1 \} \in C$ and \mbox{$ \{ 2,3,4,5\} \in M_4$} for the moment, this may be expressed as:
\begin{align}
\label{Chir}
\G_2\G_3 \G_4 \G_5  e^\pm = -e^\pm
\hspace{3cm}
\G_0\G_1   e^\pm = -e^\pm
.\end{align}
From the above relations, all information necessary to perform our desired calculations may be deduced.

It is convenient to define
\begin{align}
\label{Gpm_def}
\G_\pm =  \frac{1}{\sqrt{2}} (\G_1 \pm \G_0) 
\hspace*{0.7cm}, \hspace*{0.7cm}
\G^\pm =  \G_\mp
,\end{align}
since we as previously mentioned wish to use light-cone coordinates on the two-manifold, and to consider the action of these on the spinors instead. This may be derived in a straight-forward manner using \eqref{Gpm_def} together with \eqref{Chir}, leading to the expressions:
\begin{align}
\label{Gpm_on_epm}
\G_+ e^\pm=  \frac{1}{\sqrt{2}} (\G_1 + \G_0) e^\pm = \sqrt{2} \G_1 e^\pm
\hspace*{0.7cm}, \hspace*{0.7cm}
\G_- e^\pm=  \frac{1}{\sqrt{2}} (\G_1 - \G_0) e^\pm = 0 
.\end{align}

The most favourable way to express these relations is not however in the form in which they are given now, but rather in terms of the relations for some spinor bilinears which they lead to. Below, the most commonly used ones of these are listed:

\begin{equation}
\begin{aligned}
\label{gamma_rules}
\bar{e}^\mp \G^- e^\pm &=  1 \\
\bar{e}^\mp \G^+ e^\pm &=  0 \\
\bar{e}^\pm \G^\pm e^\pm &=  0 
\end{aligned}
\qquad
\begin{aligned}
 \bar{e}^\mp \G_\mu \G_\nu \G_+ e^\pm &= \delta_{\mu \nu} \\
  \bar{e}^\mp \G_\mu \G_\nu \G_\rho \G_\sigma \G_+ e^\pm &= \delta_{\mu \nu}\delta_{\rho \sigma}-\delta_{\mu \rho}\delta_{\nu \sigma}+\delta_{\mu \sigma}\delta_{\nu \rho}-\epsilon_{\mu \nu \rho \sigma} \\
\bar{e}^\mp \G_+ \G^- \G_+ e^\pm &=  2 
.\end{aligned}
\end{equation}

\subsection{Compactifying on $C$}

In the construction of the class $\cal{S}$ theories \cite{Gaiotto:N2dualities}, $C$ is a Riemann surface of genus $g$ with punctures. The $\mathcal{N}=2$ Yang-Mills theory arise in the IR limit of $(2,0)$ theory compactified on this surface. When considering the theory on a flat $C$, this simply means that we take all derivatives in these directions to vanish. 
Such a compactification seems necessary in our setup as well  if we want the theory on $M_4$ to be topological, since terms containing derivatives on $C$ otherwise spoil all the interesting properties of the theory: $Q$ invariance and exactness of $T^{\mu \nu}$ as well as the nilpotency of $Q$.

However, it may be interesting to point out that there are two supercharges that are Lorentz scalars on $M_4$, described by parameters $\varepsilon$ and $\bar{\varepsilon}$. We have herein chosen to only consider the observables which live in $Q$-cohomology, since we are interested in only the supercharge which would become scalar if the signature of $C$ was Euclidean and we thus could twist along that direction too.

\section{The theory after twisting}
After having worked out the field content in the previous section we now turn to the formulation of the theory after the twist. Here we will use the known equations of motion and supersymmetry variations for the abelian tensor multiplet to derive the corresponding expressions for the twisted fields. With the explicit correspondences given in section \ref{sec:fields} this is almost immediate.

\subsection{Equations of motion}

In the six dimensional formalism, the scalar fields fulfil the Klein-Gordon equation, and the self-dual bosonic three-form satisfies $dH=0$. Furthermore, the fermionic field satisfies the Dirac equation.

\begin{align}
  D^M D_M \Phi &= 0 \\
  dH &= 0 \\
  \Gamma^M D_M \Psi &= 0
\end{align}

Translated into the language of the twisted theory, the bosonic two-form and the complex scalar also satisfies the Klein-Gordon equation. Since any derivatives in the first two directions will vanish identically due to the compactification, what remains is the Klein-Gordon equation along $M_4$, that is:
\begin{align}
\label{eq:EOM_E}
\partial_\rho \partial^\rho E_{\mu \nu} &= 0 
\end{align}
\begin{align}
\label{eq:EOM_sigma}
\partial_\rho \partial^\rho \sigma &= 0
.\end{align}

Moreover, we may split the six-dimensional equation of motion for the bosonic three-form according to the number of indices along $M_4$. The six-dimensional equation of motion are then easily reinterpreted in terms of the twisted fields as:
\begin{align}
\label{eq:bos_EOM}
 2 \partial_{[\mu} A_{\nu]} &= 0 \\ \nonumber
\partial_{[\mu} F^\pm_{\nu \rho]} &= 0 \\ \nonumber
 \partial_\mu A^\mu &= 0
.\end{align}

Likewise, the equations of motion for the twisted fermionic fields may, after some calculations, be written as:
\begin{align}
\label{eq:ferm_EOM}
 \partial_\mu \tilde{\psi}^\mu 
&= 0
\\ \nonumber
\partial_\mu  \tilde{\eta}
-  \partial_\nu  \tilde{\chi}_{\mu}{}^{\nu} 
&= 0
\\ \nonumber
  (\partial_\mu  \tilde{\psi}_\nu)^+
&= 0 
,\end{align}
and equivalently for the fields without twiddles. The notation $ (\partial_\mu  \tilde{\psi}_\nu)^+$ refers to the self-dual part of $\partial_{[\mu}  \tilde{\psi}_{\nu]}$.
Furthermore, since all components of the six-dimensional fermions satisfy the Klein-Gordon equation, one can show that the same applies to all components of our twisted fermionic fields (and, as for the scalars, particularly along $M_4$).

\subsection{Supersymmetry}
After the twisting procedure we are left with two supercharges which are Lorentz scalars on $M_4$, and as explained in section \ref{sec:twist}, the one with positive $U(1)_R$ charge is the one which we focus on herein. We now derive the component expressions for this supercharge acting on the twisted fields starting from the six-dimensional expressions.  
In a flat space-time, these supersymmetry variations for the free tensor multiplet are given by:

\begin{align}
  \delta H_{M N P} &= 3\partial_{[M} \left( \bar{\Psi}_{\alpha} \Gamma_{N P]} \epsilon^\alpha \right) \label{eq:susyH}\\
  \delta \Phi_K &= 2(\Gamma^R_K)_{\alpha\beta}\bar{\Psi}^{\alpha}\epsilon^{\beta} \label{eq:susyPhi} \\ 
 \iffalse \delta \Psi^\alpha &= \frac{i}{12} H_{MNP}\Gamma^{MNP}\epsilon^\alpha + 2i M_{\beta \gamma} \partial_M \Phi^{\alpha\beta}\Gamma^M \epsilon^\gamma. \label{eq:psi_susy6} \fi
 \delta \Psi^\alpha &= \frac{i}{12} H_{MNP}\Gamma^{MNP}\epsilon^\alpha + i M_{\beta \gamma} \partial_M (\Gamma^R_K)^{\alpha \beta}\Phi^K\Gamma^M \epsilon^\gamma.\label{eq:psi_susy6}
\end{align}
Where $K$ denotes an index in the vector representation of the $R$-symmetry group $Spin(5)$.
Using the twisted field content of definitions \eqref{eq:phi_twist1}-\eqref{ferm_def} together with the supersymmetry parameter $\varepsilon$ of \eqref{SUSY} , these variations induce the following variations of the twisted fields:
\begin{equation}
\begin{aligned}
 \delta \sigma & = \sqrt{2} \tilde{\eta} v \\
\delta \bar{\sigma} & =  0 \\
  \delta E_{\mu  \nu} & =  i  \chi_{\mu  \nu} v
 \\
  \delta F_{\mu \nu}^+ & =  0 \\
  \delta  F_{\mu \nu}^- & =   - 4
  \partial_{\left[ \mu \right.} \psi_{\left. \nu \right]} v\\
  \delta  A_{\mu} & =   \partial_{\mu} \eta
   v \\
\end{aligned}
\qquad
\begin{aligned}
\delta \eta &=  
 0\\  
\delta \psi_\nu&=
 - v  i \sqrt{2}\partial_\nu \bar{\sigma}  
\\  
\delta \chi_{\mu \nu}
&= 0\\ 
\delta \tilde{\eta} &=  
 0\\  
\delta \tilde{\psi}_\nu &= i v  A_\nu-  v   \partial_\mu E_{\nu \mu}  
\\  
\delta \tilde{\chi}_{\mu \nu}
&= 2 i v F_{ \mu \nu}^+ 
\end{aligned}
\label{eq:twistedsusy}
\end{equation}
These can be verified to square to zero, which is equivalent to the supercharge $Q$ considered here indeed being nilpotent. Furthermore, these variations can be shown to induce an isomorphism on the space of solutions to the equations of motions presented in equations \eqref{eq:EOM_E}, \eqref{eq:EOM_sigma}, \eqref{eq:bos_EOM} and \eqref{eq:ferm_EOM}. 

\section{Stress tensor}
A first step towards computing the stress tensor for the theory in a general background  it is to first perform the calculations in the special case when $M_4$ has vanishing curvature. This is the subject of this section, and is something that will greatly facilitate the investigation of the general case (performed in section 5).

\subsection{Actions}
Since the main objective of this paper is to obtain an explicit expression for the stress tensor of the twisted theory, it would be highly convenient if we could formulate an action for it. The derivation of the desired stress tensor would in principle then be straight forward, and could be carried out by a standard metric variation of this action. However, as previously mentioned on repeated occasions, there are some well-known problems with giving a satisfactory formulation of $(2,0)$ theory in general, and using a Lagrangian formalism in particular, and we cannot hope to do this here either. However, there is a well-defined action for both the fermionic fields as well as the scalar fields of the abelian $(2,0)$ theory, and by writing these down we may find an Ansatz for the contributions to the stress tensor which arise from these fields. 

\subsubsection*{Scalars}
The action for the scalar field in six dimensions is given by the standard expression
\begin{align}
\mathcal{L}_{\text{scalars}} &= 
   -\partial_M\Phi^K\partial^M\Phi_K 
.\end{align}

By exploiting the fact that all derivatives in the $\pm$-directions vanish, together with the relations:
\begin{align}
\Phi_i \Phi^i &= \frac{1}{4} E_{\mu \nu} E^{\mu \nu} \\ \nonumber
\Phi_4\Phi^4+\Phi_5\Phi^5 &= 2 \sigma \bar{\sigma}
,\end{align}
the action for the scalar fields in the twisted theory may be written as:
\begin{align}
\label{eq:scalars_action}
\mathcal{L}_{\text{scalars}} &=  -\frac{1}{4}\partial_\rho E_{\mu \nu} \partial^\rho E^{\mu \nu} - 2  \partial_\rho  \sigma \partial^\rho \bar{\sigma} 
.\end{align}

\subsubsection*{Fermions}

In six dimensions, the fermionic part of the action may be written on the well-known form
\beq
\mathcal{L}=\frac{i}{2}\bar{\Psi} \G^M D_M \Psi
.\eeq
Recall that these six-dimensional fields may be reinterpreted in terms of the twisted ones according to equation \eqref{ferm_def}, which states:
\beq
\Psi= (\eta+\G_+\G_\mu \psi^\mu+\frac{1}{4}\G_{\mu \nu} \chi^{\mu \nu}) e^+ +  (\tilde{\eta}+\G_+\G_\mu \tilde{\psi}^\mu+\frac{1}{4} \G_{\mu \nu} \tilde{\chi}^{\mu \nu}) e^-
,\eeq
where $e^+$ and $e^-$ as previously are constant spinors which span the two chiral spinor representations that are Lorentz scalars on $M_4$. From this, an expression for $\bar{\Psi}$ may be obtained as:
\beq
\label{psi_bar_def}
\bar{\Psi} =  \bar{e}^+(\eta-\G_+\G_\mu \psi^\mu-\frac{1}{4}\G_{\mu \nu} \chi^{\mu \nu}) + \bar{e}^-  (\tilde{\eta}-\G_+\G_\mu \tilde{\psi}^\mu-\frac{1}{4}\G_{\mu \nu} \tilde{\chi}^{\mu \nu}) 
.\eeq

By using the properties \eqref{gamma_rules} derived for the $\G$-matrices, integration by parts and the fact that all derivatives along $C$ vanish,  the six-dimensional fermionic action  may be written in terms of the twisted fields as:  
\begin{align} \label{eq:ferm_action}
\mathcal{L}_{\text{Fermions}}
=
-i 
 \left(   \right.& \left.
  \eta \partial_\mu \tilde{\psi}^\mu 
+\psi^\mu \partial_\mu  \tilde{\eta}
-   \psi_\mu \partial_\nu  \tilde{\chi}^{\mu \nu} 
+    \chi^{\mu \nu}  \partial_\mu  \tilde{\psi}^\nu
\right) 
.\end{align}

\subsection{Ansatz and modifications}

 The stress tensor in the flat case is obtained by computing the individual contributions originating from the six-dimensional bosonic three-form, the bosonic scalar and the fermions separately, whereupon the relative coefficients are fixed by requiring supersymmetry invariance. However, which to us was somewhat unintuitive, some modifications to the terms containing the bosonic self-dual two-forms are required in order to obtain an expression which is both conserved (i.e. satisfies $D_\mu T^{\mu \nu}=0$), and $Q$-closed. This final expression of $T^{\mu \nu}$ may then be shown to also be $Q$-exact as desired.

Another important feature is that since the theory has no other definition than in terms of the equations of motion, the stress tensor will only be considered on-shell.

For the fields where an action exists, an Ansatz of the stress tensor may be computed in a standard way, namely by using
\begin{align}
\label{eq:stress_def}
T^{\mu \nu}=  \frac{1}{2} g^{\mu \nu}   \mathcal{L}+ \frac{\partial  \mathcal{L}  }{\partial g_{\mu \nu}}
.\end{align}
For the part arising from the bosonic three-form however, we are forced to take a slightly different approach. We may regard the action for a non-chiral 3-form  in six dimensions, taking the familiar expression
\begin{align}
  \cal{L} &= H_{M N P}H^{M N P} 
,\end{align}
which may be used to compute a stress tensor by the recipe stated in equation \eqref{eq:stress_def}. After this is done, the condition that $H$ is self-dual in six dimensions is imposed, and thus the first term in equation \eqref{eq:stress_def} will vanish. The remaining terms on $M_4$ will, in the language of the twisted fields, be given by
\begin{align}
\label{eq:T_H}
T_H^{\mu\nu} &=    -4 A^{(\mu} A^{\nu)} - 2F^+{}^{\mu}{}_{\rho}F^-{}^{\rho\nu} - 2 F^-{}^{\mu}{}_{\rho}F^+{}^{\rho\nu} + 
2 g^{\mu \nu} A_\rho A^\rho  
.\end{align} 

For the scalars and the fermions, one arrives at the following expressions respectively
\begin{align}
\label{eq:T_scalar}
  T^{\mu \nu}_{\Phi} &= 
-g^{\mu \nu} \partial_{\rho} \sigma \partial^\rho \bar{\sigma} + 2 \partial^{(\mu} \sigma \partial^{\nu)} \bar{\sigma} 
+ \frac{1}{4}\partial^{(\mu} E_{\rho \sigma}\partial^{\nu)}E^{\rho \sigma} 
- \frac{1}{8}g^{\mu \nu} \partial_\lambda E_{\rho \sigma} \partial^\lambda E^{\rho \sigma}
,\end{align}

\begin{align}
\label{eq:T_fermion}
T^{\mu \nu}_\Psi &=
 \frac{i}{2} g^{\mu \nu}
\left(    
  \partial_\rho \eta \tilde{\psi}^\rho 
+ \partial_\rho  \tilde{\eta} \psi^\rho
\right)
- i    \left(
 \partial^{(\mu} \eta \tilde{\psi}^{\nu)} 
+ \partial^{(\mu}  \tilde{\eta} \psi^{\nu)}
\right)
\\ \nonumber &
- \frac{i}{4} g^{\mu \nu}
 \left(  
  \tilde{\chi}^{\rho \sigma}   \partial_{[\rho}  \psi_{\sigma]}
+    \chi^{\rho \sigma}  \partial_{[\rho}  \tilde{\psi}_{\sigma]}
\right)
\\ \nonumber &
+\frac{i}{2} \left(
\chi^{\sigma (\mu}\partial_\sigma \tilde{\psi}^{\nu)} 
+\tilde{\chi}^{\sigma (\mu}\partial_\sigma \psi^{\nu)} 
-
\chi^{\sigma (\mu } \partial^{\nu)} \tilde{\psi}_{\sigma}
-\tilde{\chi}^{ \sigma (\mu  } \partial^{\nu)} \psi_{\sigma}
\right)
.\end{align}

It should be noted here that since we have self-dual fields, the variation of the metric is not as straight-forward as it would appear to be in equation \eqref{eq:stress_def}. This is because the condition of self-duality contains an implicit metric dependence, and thus a variation of the metric must be accompanied by a variation of all self-dual fields present. A term consisting of such a self-dual field, $\chi_{\mu \nu}$, with indices contracted with some other rank-2 tensor, $X^{\mu \nu}$, will under a metric variation take the form:

\begin{align}
  X^{\mu \nu}\delta_g \chi_{\mu \nu} &= 
 -\frac{1}{4}  \delta g_{\mu \nu}  g^{\mu \nu}  X_{\kappa \lambda}   \chi^{\kappa \lambda}
 +\delta g_{\mu \nu}  X^{[\mu \sigma]}  \chi^{\nu}{}_{ \sigma}
. \end{align}

The three pieces in \eqref{eq:T_H}, \eqref{eq:T_scalar} and \eqref{eq:T_fermion} are each conserved individually, which may be shown by straight-forward, but yet tedious calculations that are omitted here.
In order to stand a chance of fulfilling supersymmetry invariance under the transformations listed in equation \eqref{eq:twistedsusy}, the relative coefficients amongst the different contributions are fixed. The stress tensor one then finds is given by:

\begin{align}
  T^{\mu \nu} &= \frac{1}{2} \Big(
-g^{\mu \nu} \partial_{\rho} \sigma \partial^\rho \bar{\sigma} + 2 \partial^{(\mu} \sigma \partial^{\nu)} \bar{\sigma} 
+ \frac{1}{4}\partial^{(\mu} E_{\rho \sigma}\partial^{\nu)}E^{\rho \sigma} 
- \frac{1}{8}g^{\mu \nu} \partial_\lambda E_{\rho \sigma} \partial^\lambda E^{\rho \sigma}
\Big)
\\  \nonumber
& +\frac{1}{8}\left( -4 A^{(\mu} A^{\nu)} - 2F^+{}^{\mu}{}_{\rho}F^-{}^{\rho\nu} - 2 F^-{}^{\mu}{}_{\rho}F^+{}^{\rho\nu} + 
2 g^{\mu \nu} A_\rho A^\rho  \right)
\\ \nonumber
&+ 
 \frac{i}{2} g^{\mu \nu}
\left(    
  \partial_\rho \eta \tilde{\psi}^\rho 
+ \partial_\rho  \tilde{\eta} \psi^\rho
\right)
- i    \left(
 \partial^{(\mu} \eta \tilde{\psi}^{\nu)} 
+ \partial^{(\mu}  \tilde{\eta} \psi^{\nu)}
\right)
\\ \nonumber &
- \frac{i}{4} g^{\mu \nu}
 \left(  
  \tilde{\chi}^{\rho \sigma}   \partial_{[\rho}  \psi_{\sigma]}
+    \chi^{\rho \sigma}  \partial_{[\rho}  \tilde{\psi}_{\sigma]}
\right)
\\ \nonumber &
+\frac{i}{2} \left(
\chi^{\sigma (\mu}\partial_\sigma \tilde{\psi}^{\nu)} 
+\tilde{\chi}^{\sigma (\mu}\partial_\sigma \psi^{\nu)} 
-
\chi^{\sigma (\mu } \partial^{\nu)} \tilde{\psi}_{\sigma}
-\tilde{\chi}^{ \sigma (\mu  } \partial^{\nu)} \psi_{\sigma}
\right) 
.\end{align}

However, some problematic terms still exist which prevents the above expression from being  $Q$-closed. By a long and quite intricate calculation, one may show that this obstruction is solved if the part of the stress tensor containing the self-dual two-form which arose from the six-dimensional scalars, namely terms containing $E_{\mu \nu}$, is altered to:
\begin{align}
 T^{\mu \nu}_{EE\text{-terms}} &=
\frac{1}{4}g^{\mu \nu}     \partial^\kappa E_{\rho \kappa}    \partial_\sigma E^{\rho \sigma}
-  \frac{1}{2}   \partial^\rho \Big(  \partial^\kappa E^{(\mu}{}_{ \kappa}  E^{\nu)}{}_{\rho} \Big)
+ \frac{1}{2}  \partial^{(\mu}    \partial_\kappa E^{\rho \kappa}  E^{\nu)}{}_{\rho} 
.\end{align}
Also this part is conserved on its own, and so this alteration preserves the conservation of $T^{\mu \nu}$. This may be shown by a slightly more complicated calculation than for any of the other terms, which requires the repeated use of the self-duality of $E_{\mu \nu}$.

That this problem of supersymmetry invariance is solved by altering the terms containing the fields originating from the scalars, for which we had an action from which to derive a stress tensor, may seem quite unintuitive. However, we must bear in mind that even though we have an action for some fields in the theory, there is no action for the \emph{entire} theory. Hence we do not have a supersymmetric quantity from which we may derive a supersymmetric stress tensor, and though using the actions presented in equations \eqref{eq:ferm_action} and \eqref{eq:scalars_action} provides us with a good Ansatz for a stress tensor for the entire theory, we should not expect this approach to give us a supersymmetric result. 

 The complete stress tensor for this theory when placed on a flat background may then finally be written down explicitly as 

\begin{align}
\label{eq:Stress_Tensor_flat}
  T^{\mu \nu} &= \frac{1}{2} \Big(
-g^{\mu \nu} \partial_{\rho} \sigma \partial^\rho \bar{\sigma} + 2 \partial^{(\mu} \sigma \partial^{\nu)} \bar{\sigma} 
\Big)
\\  \nonumber
& +\frac{1}{8}\left( -4 A^{(\mu} A^{\nu)} - 2F^+{}^{\mu}{}_{\rho}F^-{}^{\rho\nu} - 2 F^-{}^{\mu}{}_{\rho}F^+{}^{\rho\nu} + 
2 g^{\mu \nu} A_\rho A^\rho  \right)
\\ \nonumber
&+ 
 \frac{i}{2} g^{\mu \nu}
\left(    
  \partial_\rho \eta \tilde{\psi}^\rho 
+ \partial_\rho  \tilde{\eta} \psi^\rho
\right)
- i    \left(
 \partial^{(\mu} \eta \tilde{\psi}^{\nu)} 
+ \partial^{(\mu}  \tilde{\eta} \psi^{\nu)}
\right)
\\ \nonumber &
- \frac{i}{4} g^{\mu \nu}
 \left(  
  \tilde{\chi}^{\rho \sigma}   \partial_{[\rho}  \psi_{\sigma]}
+    \chi^{\rho \sigma}  \partial_{[\rho}  \tilde{\psi}_{\sigma]}
\right)
\\ \nonumber &
+\frac{i}{2} \left(
\chi^{\sigma (\mu}\partial_\sigma \tilde{\psi}^{\nu)} 
+\tilde{\chi}^{\sigma (\mu}\partial_\sigma \psi^{\nu)} 
-
\chi^{\sigma (\mu } \partial^{\nu)} \tilde{\psi}_{\sigma}
-\tilde{\chi}^{ \sigma (\mu  } \partial^{\nu)} \psi_{\sigma}
\right) 
\\ \nonumber
&
+ \frac{1}{4}g^{\mu \nu}     \partial^\kappa E_{\rho \kappa}    \partial_\sigma E^{\rho \sigma}
-  \frac{1}{2}   \partial^\rho \Big(  \partial^\kappa E^{(\mu}{}_{ \kappa}  E^{\nu)}{}_{\rho} \Big)
+ \frac{1}{2}  \partial^{(\mu}    \partial_\kappa E^{\rho \kappa}  E^{\nu)}{}_{\rho} 
,\end{align}
where the last line above is the manually altered terms that are needed to make the stress tensor invariant under the supersymmetry transformations in equation \eqref{eq:twistedsusy}.

\subsection{$Q$-exactness}
The stress tensor presented above in  \eqref{eq:Stress_Tensor_flat}  is after an examination found to be  $Q$-exact and may be written as

\begin{align}
  T^{\mu \nu} &=  \Big\{  Q, \lambda^{\mu \nu}  \Big\}
,\end{align}
where
\begin{align}
\label{eq:lambda}
  \lambda^{\mu \nu} &=
  \frac{1}{2}\Big(
 \sqrt{2}i \psi^{(\mu} \partial^{\nu)} \sigma 
 + \tilde{\psi}^{(\mu} \partial^\rho E^{\nu)}{}_{\rho} 
 + \partial_{\rho}\tilde{\psi}^{(\mu} E^{\nu)}{}_{\rho} 
 - \partial^{(\mu}\tilde{\psi}^{\rho}  E^{\nu)}{}_{\rho} 
\\ \nonumber &\phantom{=} 
+ i \tilde{\psi}^{(\mu} A^{\nu)}
- \frac{i}{2}  \tilde{\chi}^{(\mu}{}_\rho  F^-{}^{ \nu)}{}^{\rho} 
-\frac{i}{\sqrt{2} }g^{\mu \nu}\psi_\rho \partial^\rho \sigma
-\frac{1}{2}g^{\mu \nu}\tilde{\psi_\rho} \partial_\sigma E_{\rho \sigma}
- \frac{i}{2} g^{\mu \nu} \tilde{\psi}_\rho A^\rho \Big)
.\end{align}

To find $\lambda^{\mu \nu}$, an Ansatz was used in which all possible allowed, terms were included. These are however not as many as one may think, since there are constraints due to dimensionality and $U(1)$-charge. These constraints forces us to restrict ourselves to terms of dimensionality $11/2$ and $U(1)$ charge of $-1/2$, (which all of the above terms clearly satisfy). In table \ref{tab:dim}, the dimensionality and $U(1)$-charge of the different fields, as well as the supersymmetry parameter and stress tensor, are listed. 

\begin{table}[hbt]
  \[
\begin{array}{c | c | c}
 & \text{dimensionality} & U(1)_R\text{-charge  }\\
\hline
\eta, \psi_\mu, \chi_{\mu \nu} & 5/2 & +1/2  \\
\tilde{\eta},  \tilde{\psi}_\mu,  \tilde{\chi}_{\mu \nu} & 5/2 & -1/2 \\
\bar{\sigma} & 2 & +1 \\
\sigma & 2 & -1 \\
E_{\mu \nu} & 2 & 0  \\
A_\mu, F_{\mu \nu} & 3 & 0  \\
T_{\mu \nu} & 6 & 0  \\
\epsilon & -1/2& -1/2 \\
\end{array} 
\]
\caption{Mass dimension and $U(1)_R$ charges of the fields, parameters and curvature tensors.}
\label{tab:dim}
\end{table}

\section{The case when $M_4$ is curved}
In the previous section, an expression for the stress tensor when $M_4$  has vanishing curvature is obtained and shown to indeed be $Q$-exact. This was done by explicitly finding a $\lambda^{\mu \nu}$ such that $T^{\mu \nu}= \{ Q, \lambda^{\mu \nu} \}$.  Now we are faced with the question:  How does this change in the case when $M_4$ is curved?

A simple starting point here would instead be to ask the question ``How may $\lambda^{\mu \nu}$ change when $M_4$ becomes curved?''.  The restrictions imposed upon $\lambda^{\mu \nu}$ by dimensionality may be used here as well.  Since $\lambda^{\mu \nu}$ is of fractional dimension, an odd number of fermionic fields must be included. Also, since we wish to add terms related to curvature, the Riemann-, Ricci-tensor or curvature scalar must be included in these, each of which is of dimension $2$. The remaining part of these terms must be of dimension $1$, which means that our only option is to incorporate a derivative. Terms like these are however not bilinears in the fields, and thus make no sense at all. 

By the reasoning above, there are no terms which may possibly be added to $\lambda^{\mu \nu}$ in the case when $M_4$ is curved. Thus, the stress tensor even in this case will still be given by the expression $\{ Q, \lambda^{\mu \nu}\}$.

It should be noted that there are two more places that could be modified in the curved case:  the scalar equations of motion and the fermion supersymmetry variations.

The scalar equations of motion could be modified to replace the right hand side of the Klein-Gordon equation in both  \eqref{eq:EOM_E} and \eqref{eq:EOM_sigma} with a multiple of the curvature scalar multiplying the fields. However, such a modification in \eqref{eq:EOM_sigma} would ruin the conservation properties of the part of the stress tensor containing the bosonic scalars and is thus not allowed. The same modification in \eqref{eq:EOM_E},
\begin{align}
\label{eq:EOM_E_curved}
D_\rho D^\rho E_{\mu \nu} &= a R E_{\mu \nu},
\end{align}
may be carried out, where $a$ is some constant. However, this will not be enough to rectify the problems arising when $M_4$ is curved, something which is further discussed in section \ref{sec:conservation}.

The fermionic supersymmetry variations for the six-dimensional free tensor multiplet in a curved background may contain an extra term of the form
\begin{equation}
  \delta \Psi = \dots + \Phi \Gamma^M D_M \epsilon.
\end{equation}
This term will not contribute to the twisted supersymmetry transformations since the whole point of the twisting is to manufacture a covariantly constant supercharge.

Thus, in the curved case, the stress tensor cannot be subject to any modifications and will still  be given by $\Big\{  Q, \lambda^{\mu \nu}  \Big\}$, where all partial derivatives in $\lambda^{\mu \nu}$ are now replaced by covariant ones. This gives us $T^{\mu \nu}$ as in equation \eqref{eq:Stress_Tensor_flat} but again, with partial derivatives replaced by covariant ones. The generalisation to a curved $M_4$ is thus:
\begin{align}
\label{eq:Stress_Tensor_curved}
  T^{\mu \nu} &= \frac{1}{2} \Big(
-g^{\mu \nu} D_{\rho} \sigma D^\rho \bar{\sigma} + 2 D^{(\mu} \sigma D^{\nu)} \bar{\sigma} 
\Big)
\\  \nonumber
& +\frac{1}{8}\left( -4 A^{(\mu} A^{\nu)} - 2F^+{}^{\mu}{}_{\rho}F^-{}^{\rho\nu} - 2 F^-{}^{\mu}{}_{\rho}F^+{}^{\rho\nu} + 
2 g^{\mu \nu} A_\rho A^\rho  \right)
\\ \nonumber
&+ 
 \frac{i}{2} g^{\mu \nu}
\left(    
  D_\rho \eta \tilde{\psi}^\rho 
+ D_\rho  \tilde{\eta} \psi^\rho
\right)
- i    \left(
 D^{(\mu} \eta \tilde{\psi}^{\nu)} 
+ D^{(\mu}  \tilde{\eta} \psi^{\nu)}
\right)
\\ \nonumber &
- \frac{i}{4} g^{\mu \nu}
 \left(  
  \tilde{\chi}^{\rho \sigma}   D_{[\rho}  \psi_{\sigma]}
+    \chi^{\rho \sigma}  D_{[\rho}  \tilde{\psi}_{\sigma]}
\right)
\\ \nonumber &
+\frac{i}{2} \left(
\chi^{\sigma (\mu}D_\sigma \tilde{\psi}^{\nu)} 
+\tilde{\chi}^{\sigma (\mu}D_\sigma \psi^{\nu)} 
-
\chi^{\sigma (\mu } D^{\nu)} \tilde{\psi}_{\sigma}
-\tilde{\chi}^{ \sigma (\mu  } D^{\nu)} \psi_{\sigma}
\right) 
\\ \nonumber
&
+ \frac{1}{4}g^{\mu \nu}     D^\kappa E_{\rho \kappa}    D_\sigma E^{\rho \sigma}
-  \frac{1}{2}   D^\rho \Big(  D^\kappa E^{(\mu}{}_{\kappa}  E^{\nu)}{}_{\rho} \Big)
+ \frac{1}{2}  D^{(\mu}    D_\kappa E^{\rho \kappa}  E^{\nu)}{}_{\rho} 
.\end{align}

That this stress tensor is still $Q$-exact is obvious, but it is not completely clear that it still fulfils the criteria of being covariantly conserved. Rather surprisingly, it would seem that it does \emph{not}. Again, the complications lie in the part containing the self-dual bosonic two-forms. By considering the covariant derivative of these terms, the complications arising here for a curved $M_4$ will be apparent.

\subsection{Covariant conservation of $T^{\mu \nu}$ in the curved case \label{sec:conservation}} 
Consider the covariant divergence of the terms containing the bosonic self-dual two-forms:
\begin{align}
\label{eq:T_EE}
D^\mu T_{\mu \nu}^{\text{curved }EE\text{-terms}}  =&
+ \frac{1}{2}g_{\mu \nu}   D^\mu D^\kappa E_{\rho \kappa}   D_\sigma E^{\rho \sigma}
-  \frac{1}{2}  D^{[\mu} D^{\rho]} \Big( D_\kappa E_{(\mu}{}^{ \kappa}  E_{\nu) \rho} \Big)
\\ \nonumber &
-  \frac{1}{2}  D^{(\mu} D^{\rho)} \Big( D_\kappa E_{(\mu}{}^{ \kappa}  E_{\nu) \rho} \Big)
+ \frac{1}{2} D^\mu \Big( D_{(\mu}   D_\kappa E^{\rho \kappa}  E_{\nu) \rho} \Big)
.\end{align}

This can, as previously mentioned, be shown to vanish when $M_4$ is flat, but  in the curved case, there are additional terms arising from commuting the derivatives which may yet cause problems. A few of the above terms will give rise to terms containing derivatives on the curvature tensors, which must cancel on their own for any chance to maintain conservation of $T^{\mu \nu}$. Such terms will arise from terms containing  three derivatives acting on the same field, that is from the two last terms in the expression above.

Let us start by considering terms of this kind. By using two forms of the Bianchi identity,  together with a basis expansion of the self-dual two-forms according to $E_{\mu \nu} = E_i T^{i}_{\mu \nu}$, (where $i \in \{ 1,2,3\}$ and the $T^i$'s form a basis on the space of self-dual two-forms) in the cases where the two bosonic fields are contracted, one may in a straight-forward manner show that all terms containing the derivatives on the curvature tensors may be written as:
\beq
\label{eq:der_curve3}
- \frac{1}{4}    D_{\tau} R_{\rho \kappa  } E^{\tau \kappa}  E_{\nu}{}^{\rho}
+ \frac{1}{8} D_{\nu}  R_{\mu \kappa \rho \tau}  E^{\tau \kappa}  E^{\mu \rho}
+ \frac{1-2a}{4} D_{\nu}  R 
 E_i  E^i
.\eeq
To obtain this expression, the most general form of the equations of motions for $E_{\mu\nu}$ on a curved background were used, as given in \eqref{eq:EOM_E_curved}.

This is in general non-zero, which may be easily shown by introducing a concrete example in which this quantity does not vanish. An example of such a configuration is $M_4 = \mathbb{R} \times M_3$, where index value $1$ denotes the coordinate along $\mathbb{R}$, and $M_3$ is of non-vanishing curvature. Consider (\ref{eq:der_curve3}) in the case where $\nu=1$. In such a case, the two last terms vanish, where as the first one in general does not.
 We have thus shown that the unique, $Q$-exact stress tensor of the theory is not conserved when the theory is placed upon a general four-manifold $M_4$.
 
 That the divergence of the suggested stress tensor is non-vanishing is deeply unsettling since in general, conservation of the stress tensor is a direct consequence of general covariance of the theory. In particular, any stress tensor for a topological field theory needs to satisfy this, as well as being $Q$-exact\footnote{This is a characteristic feature of the class of topological field theories considered herein.}. Our result thus implies that the theory obtained herein does not appear to be topological. 
  One should however notice that no claims are made as to the situation when the condition of $Q$-exactness of $T^{\mu \nu}$ is removed. Such a setup lies outside the scope of this work.

\section{Conclusion and outlook}
Herein, we have shown that there is no possible covariantly conserved, $Q$-exact stress tensor when this twisted form of the theory is placed on a general background. The twisting in question is taken to be the one described, for example, in \cite{Yagi:1} where the free tensor multiplet of $(2,0)$ theory is placed on $M_6=C \times M_4$ with Minkowski signature. Twisting along $M_4$  and  compactification along $C$ as described in section 2 is then performed. Furthermore, the theory is only considered under the supercharge that would become scalar on $C$ \emph{if} it were of Euclidean signature and further twistings could be performed. This must however remain an ``\emph{if}'', because of the problems surrounding the formulation of $(2,0)$ theory, especially in Euclidean signature. Because of these problems, all of our investigations were kept on the level of equations of motion.

The lack of a $Q$-exact stress tensor in general is to us a surprising result, and we shall below discuss some potential explanations to its origin:

One obvious possible source of this unexpected result may be found in the details of the twisting procedure itself. It should be noted that when compactifying the $(2,0)$ theory on a general two-manifold, $C$, all supersymmetry is broken and there is thus no way to twist the resulting four-dimensional theory. However, when the six-dimensional theory is twisted (in the Euclidean case), one supercharge is preserved, and thus a successive compactification on $C$ could result in a supersymmetric theory. The moral of this is that in general, a twist of the six-dimensional theory cannot be seen as a twisting obtained through a four-dimensional super Yang-Mills theory.

A more explicit way to illustrate this is through the following rather simple scenario:  Consider the free $(2,0)$ theory on a Euclidean six-manifold with a  product structure, $C \times M_4$. The simplest possible approach to obtain a four-dimensional supersymmetric theory would be to take $C$ to be a flat two-torus and compactify on this to obtain the well-known $\mathcal{N}=4$ Yang-Mills theory. This theory in turn admits three \emph{inequivalent} topological twists \cite{VafaWitten}, resulting in three different four-dimensional theories.
However, starting at the other end, the six-dimensional $(2,0)$ theory admits one \emph{unique} twist. After this is carried out, one may compactify on the two-torus and obtain some particular four-dimensional theory.
Hence these two approaches cannot be considered equivalent.

The conjecture that the twisted theory along $M_4$ should coincide with Donaldson-Witten theory 
is obtained through thinking about the \linebreak six-dimensional topological twist in the manner discussed in the above paragraphs. That is, it is considered equivalent to first compactfying down to the class $\mathcal{S}$ theories and then twisting these. Our arguments above, together with the obtained result of section \ref{sec:conservation} would seem to indicate that the situation may actually be more complicated.

However, the reasoning above is in full carried out in a Euclidean scenario where, as noted on many occasions, no satisfactory formulation of the (2,0) theory exist. Rather all the calculations presented in this paper are carried out in a Lorenzian signature, and the twisting along $C$ is for obvious reasons not carried out. One could then ask if this situation finds its remedy in the hypothetical twisting along $C$.  This will however not be the case since this twisting would only result in different $U(1)$-charges of the fields, and all arguments done here for possible curvature corrections etc are not dependent on this, but rather on dimensionality which remains unchanged.
Moreover, it shall be noted that the issue of signature also affects the compactification on the two-manifold, which may be a contributing factor to the obtained result.

Another possible resolution of these difficulties may be found in a hypothetical formulation of the free $(2,0)$ tensor multiplet in a Euclidean signature. If one requires that this hypothetical theory should indeed give rise to a topological field theory under the twisting described herein, this investigation of the difficulties presented for its Minkowski analog may shed some light on desired properties of the Euclidean theory.

\pagebreak

\section*{Acknowledgement}
We would like to express our gratitude to Måns Henningson for suggesting the problem and the continued support throughout this project. The authors also wish to thank Martin Cederwall for useful discussions. \\ \\
This research was supported by the Swedish Research Council (VR).

\bibliography{refs}{}

\providecommand{\href}[2]{#2}\begingroup\raggedright\begin{thebibliography}{10}

\bibitem{Witten:1995}
E.~Witten, {\it {Some comments on string dynamics}},
  \href{http://xxx.lanl.gov/abs/hep-th/9507121}{{\tt hep-th/9507121}}.

\bibitem{Witten:ct46}
E.~Witten, {\it {Conformal Field Theory In Four And Six Dimensions}},
  \href{http://xxx.lanl.gov/abs/0712.0157}{{\tt arXiv:0712.0157}}.

\bibitem{Seiberg:noteon16}
N.~Seiberg, {\it {Notes on theories with 16 supercharges}},  {\em
  Nucl.Phys.Proc.Suppl.} {\bf 67} (1998) 158--171,
  [\href{http://xxx.lanl.gov/abs/hep-th/9705117}{{\tt hep-th/9705117}}].

\bibitem{Gaiotto:N2dualities}
D.~Gaiotto, {\it {N=2 dualities}},  {\em JHEP} {\bf 1208} (2012) 034,
  [\href{http://xxx.lanl.gov/abs/0904.2715}{{\tt arXiv:0904.2715}}].

\bibitem{Klemm:sclass}
A.~Klemm, W.~Lerche, P.~Mayr, C.~Vafa, and N.~P. Warner, {\it {Selfdual strings
  and N=2 supersymmetric field theory}},  {\em Nucl.Phys.} {\bf B477} (1996)
  746--766, [\href{http://xxx.lanl.gov/abs/hep-th/9604034}{{\tt
  hep-th/9604034}}].

\bibitem{Witten:sclass}
E.~Witten, {\it {Solutions of four-dimensional field theories via M theory}},
  {\em Nucl.Phys.} {\bf B500} (1997) 3--42,
  [\href{http://xxx.lanl.gov/abs/hep-th/9703166}{{\tt hep-th/9703166}}].

\bibitem{Gaiotto:2009}
D.~Gaiotto, G.~W. Moore, and A.~Neitzke, {\it {Wall-crossing, Hitchin Systems,
  and the WKB Approximation}},  \href{http://xxx.lanl.gov/abs/0907.3987}{{\tt
  arXiv:0907.3987}}.

\bibitem{Gadde:2013}
A.~Gadde, S.~Gukov, and P.~Putrov, {\it {Fivebranes and 4-manifolds}},
  \href{http://xxx.lanl.gov/abs/1306.4320}{{\tt arXiv:1306.4320}}.

\bibitem{AGT}
L.~F. Alday, D.~Gaiotto, and Y.~Tachikawa, {\it {Liouville Correlation
  Functions from Four-dimensional Gauge Theories}},  {\em Lett.Math.Phys.} {\bf
  91} (2010) 167--197, [\href{http://xxx.lanl.gov/abs/0906.3219}{{\tt
  arXiv:0906.3219}}].

\bibitem{Nekrasov:2002}
N.~A. Nekrasov, {\it {Seiberg-Witten prepotential from instanton counting}},
  {\em Adv.Theor.Math.Phys.} {\bf 7} (2004) 831--864,
  [\href{http://xxx.lanl.gov/abs/hep-th/0206161}{{\tt hep-th/0206161}}].

\bibitem{Nekrasov:2003}
N.~Nekrasov and A.~Okounkov, {\it {Seiberg-Witten theory and random
  partitions}},  \href{http://xxx.lanl.gov/abs/hep-th/0306238}{{\tt
  hep-th/0306238}}.

\bibitem{Yagi:1}
J.~Yagi, {\it {On the Six-Dimensional Origin of the AGT Correspondence}},  {\em
  JHEP} {\bf 1202} (2012) 020, [\href{http://xxx.lanl.gov/abs/1112.0260}{{\tt
  arXiv:1112.0260}}].

\bibitem{Yagi:2}
J.~Yagi, {\it {Compactification on the $\Omega$-background and the AGT
  correspondence}},  {\em JHEP} {\bf 1209} (2012) 101,
  [\href{http://xxx.lanl.gov/abs/1205.6820}{{\tt arXiv:1205.6820}}].

\bibitem{Witten:1988}
E.~Witten, {\it {Topological Quantum Field Theory}},  {\em Commun.Math.Phys.}
  {\bf 117} (1988) 353.

\bibitem{Witten:2011}
E.~Witten, {\it {Fivebranes and Knots}},
  \href{http://xxx.lanl.gov/abs/1101.3216}{{\tt arXiv:1101.3216}}.

\bibitem{Silverstein:Witten}
E.~Silverstein and E.~Witten, {\it {Global U(1) R symmetry and conformal
  invariance of (0,2) models}},  {\em Phys.Lett.} {\bf B328} (1994) 307--311,
  [\href{http://xxx.lanl.gov/abs/hep-th/9403054}{{\tt hep-th/9403054}}].

\bibitem{Howe:1983}
P.~S. Howe, G.~Sierra, and P.~Townsend, {\it {Supersymmetry in
  Six-Dimensions}},  {\em Nucl.Phys.} {\bf B221} (1983) 331.

\bibitem{VafaWitten}
C.~Vafa and E.~Witten, {\it {A Strong coupling test of S duality}},  {\em
  Nucl.Phys.} {\bf B431} (1994) 3--77,
  [\href{http://xxx.lanl.gov/abs/hep-th/9408074}{{\tt hep-th/9408074}}].

\end{thebibliography}\endgroup
\bibliographystyle{JHEP}

\end{document}